\documentclass{aa}  

\usepackage{graphicx}
\usepackage{txfonts}
\usepackage[colorlinks,citecolor=blue]{hyperref}
\usepackage[rgb]{xcolor}
\usepackage[normalem]{ulem}

\usepackage{orcidlink}

\begin{document}

\title{Composition-asymmetric and sheared relativistic magnetic reconnection}


   \author{Enzo Figueiredo\inst{1}\orcidlink{0009-0006-7103-1965}
          \and
          Beno\^it Cerutti\inst{1}\orcidlink{0000-0001-6295-596X}
          \and
          John Mehlhaff\inst{1}\orcidlink{0000-0002-7414-0175}
          \and
          Nicolas Scepi\inst{1}\orcidlink{0000-0003-3909-2486}
          }

\institute{Univ. Grenoble Alpes, CNRS, IPAG, 38000 Grenoble, France\\
           \email{enzo.figueiredo@univ-grenoble-alpes.fr}
           }

\date{Received \today; accepted ...}

 
  \abstract
   {Relativistic magnetic reconnection studies have focused on symmetric configurations so far, where the upstream plasma has identical properties on each side of the layer. Yet, just like nonrelativistic reconnection on the day side of the Earth's magnetosphere, relativistic reconnection may also operate at the interface between highly asymmetric environments. The boundary layer between a relativistic jet and an accretion flow forming around a supermassive black hole may present such asymmetric configurations in terms of plasma composition, bulk velocity, temperature and magnetization.}
   {In this work, we aim to conduct the first study of relativistic magnetic reconnection where the upstream plasma is composed of electron-positron pairs on one side, and electrons and ions on the other. We also investigate the role of a relativistic symmetric shear flow applied along the reconnecting field lines.}
   {We simulate magnetic reconnection using two-dimensional particle-in-cell simulations. The initial setup is adapted from a classic Harris layer without guide field, modified to accommodate plasma-composition and shear asymmetries in the upstream medium.}
   {For a composition-asymmetric setup, we find that the reconnection dynamics is driven by the electron-ion side, which is the plasma with the lowest magnetization. The energy partition favors accelerating ions at the expense of electrons even more than in a corresponding symmetric setup. With respect to shear, a super-Alfv\'enic upstream decreases the laboratory-frame reconnection rate, but, unlike in non-relativistic studies, does not shut off reconnection completely.}
   {The asymmetries examined in this work diminish the overall efficiency of electron acceleration relative to corresponding symmetric configurations. In the context of a black hole jet-disk boundary, asymmetric reconnection alone is probably not efficient at accelerating electrons to very high energies, but it might facilitate plasma mixing and particle injection for other acceleration channels at the interface.}


   \maketitle
%
\section{Introduction}

Magnetic reconnection is a widely studied plasma phenomenon, heating and accelerating particles through a change in the global magnetic topology \citep{Zweibel2009}. This process typically occurs when two antiparallel magnetic field lines snap and recombine across a thin current layer. The relativistic regime of reconnection has encountered wider interest in the last decade, as it provides efficient nonthermal particle acceleration (e.g., \citealt{Hoshino2001, Hoshino2012, 2012ApJ...754L..33C, Sironi2014, 2014PhRvL.113o5005G, Kagan2015, 2016ApJ...816L...8W}). Here, in contrast to non-relativistic reconnection, the energy associated with the magnetic field dominates over the plasma rest mass energy, providing a significant free-energy reservoir for particle energization. Several regimes of relativistic magnetic reconnection have now been studied, including: with different plasma compositions (i.e., electron-positron pairs, and electron-ion plasmas, \citealt{Melzani2014b, Werner2018}); involving strong radiative cooling and pair production \citep{2009PhRvL.103g5002J, Cerutti2013, 2019ApJ...877...53H, Schoeffler2019, 2024MNRAS.52711587M}; and driven by other plasma processes such as turbulence \citep{2017PhRvL.118e5103Z, Comisso2018, 2023ApJ...944..122M} and Kelvin-Helmholtz or Rayleigh-Taylor instabilities \citep{2020A&A...642A.123C, Sironi2021, Zhdankin2023}.

Relativistic magnetic reconnection is a key driver of high-energy emission in various astrophysical environments. In pulsar magnetospheres, the equatorial current sheet forming beyond the light cylinder undergoes reconnection in a strongly radiative and pair-producing regime, emitting gamma-ray emission through synchrotron radiation \citep{Lyubarskii1996, Uzdensky&Spitkovsky2014, 2016MNRAS.457.2401C}. Similar processes may also power gamma-ray flares in the Crab nebula \citep{2011ApJ...737L..40U, Cerutti2012}, magnetar eruptions and fast radio bursts \citep{2003MNRAS.346..540L, 2020ApJ...900L..21Y, 2022ApJ...932L..20M}. In active galactic nuclei, magnetic reconnection is considered a plausible mechanism for explaining gamma-ray flares within blazar jets \citep{2009MNRAS.395L..29G, 2011MNRAS.413..333N, 2016MNRAS.462.3325P, 2020MNRAS.498..799M} and also bursting activity closer to the central supermassive black hole, for instance X-ray flares in SgrA* \citep{Ball2016, Scepi2022} and gamma-ray flares in M87* \citep{Crinquand2021, Ripperda2022, 2023ApJ...943L..29H,2024arXiv240601211S}.

Recent VLBI observations of M87* show a clear edge-brightened jet connected to the innermost parts of the accretion flow \citep{Lu2023}, suggesting that particles are accelerated at the interface between these two entities. In the framework of a Poynting-flux dominated jet \citep{Blandford&Znajek1977}, this boundary layer may separate a relativistic ultra-magnetized jet loaded with pairs produced near the ergosphere on one side from a hot mildly magnetized accretion flow composed of electrons and ions on the other side. Such abrupt discontinuities in plasma composition, temperature, bulk velocity and magnetization may lead to qualitatively different particle acceleration compared to a symmetric environment.

Recent global modeling efforts suggest magnetic reconnection as a prominent particle acceleration mechanism at such highly asymmetric jet-disk boundaries. The accretion of magnetic loops onto a spinning black hole leads to the formation of reconnecting current sheets between the jet and the disk  
\citep{2015MNRAS.446L..61P, 2019MNRAS.487.4114Y, 2022A&A...663A.169E, ElMellah2023}. Magnetic flux eruptions reported in general relativistic magnetohydrodynamic (GRMHD) simulations also provoke large-scale reconnection events that push magnetized flux tubes through the accretion flow (e.g., see \citealt{Igumenshchev2008, Tchekhovskoy2011}). This phenomenon can be accompanied with additional current sheet formation at the interface between the jet and the disk \citep{Ripperda2020, Ripperda2022, 2021MNRAS.508.1241C, Vos2023}. However, it is unclear how magnetic reconnection and particle acceleration proceed within such asymmetric boundary layers.
 
Asymmetric and sheared magnetic reconnection have been previously studied in the non-relativistic regime, in particular in the context of the interaction between the Earth's magnetosphere and the solar wind (i.e., the magnetopause). An important takeaway message from these studies is that a strong velocity shear applied along the field lines decreases the reconnection rate. It can even halt the process entirely for super-Alfv\'enic flows \citep{LaBelle-Hamer1995, Cassak2011a, Cassak2011b}. An asymmetry in the upstream plasma density and magnetic field strength leads to a global reorganization of the current layer and a change in the plasmoids' shape \citep{Murphy2012, Eastwood2013}. However, from these studies alone it is unclear how such asymmetries would affect particle acceleration in a relativistic context.

The first study of relativistic asymmetric magnetic reconnection was recently carried out by \citet{Mbarek2022}. They simulate a pair plasma with an upstream density and magnetic field strength contrast on each side of the reconnection layer, hence creating a magnetization asymmetry in the reconnection process. They find that the lowest magnetization side dictates the reconnection rate, and that particle acceleration remains in an intermediate regime between the higher and lower magnetizations. This result suggests that particle acceleration via reconnection could be quenched at a black hole jet-disk interface. 

In this study, our first objective is to investigate the role of plasma asymmetries on particle acceleration and mixing within relativistic reconnection layers, using two-dimensional (2D) particle-in-cell (PIC) simulations. More specifically, and in contrast to \citet{Mbarek2022}, we investigate the role of a plasma-composition asymmetry where one side is filled with an ultra-relativistic ultra-magnetized plasma of pairs, while the other side is composed of a mildly relativistic and magnetized electron-ion plasma. This configuration is reminiscent of a black hole jet-disk boundary layer. 
Our second objective is to evaluate the impact of a relativistic shear flow directed along the field lines -- another property reminiscent of the jet-disk boundary -- on the reconnection rate and on the particle acceleration efficiency.

This article is organized as follows. In Section~\ref{sec:Setup}, we detail the modifications made to the classic relativistic Harris setup to implement a velocity shear and a composition asymmetry in the upstream medium. In sections~\ref{sec:SpecAsym} and \ref{sec:Shear}, we respectively present our findings on composition asymmetry and shear. We summarize and discuss the implications of our results in Section~\ref{sec:Conclusion}.

\section{Numerical setup} \label{sec:Setup}

\subsection{Asymmetric Harris equilibrium}

\begin{figure}[h]
    \centering
    \includegraphics[width=\columnwidth]{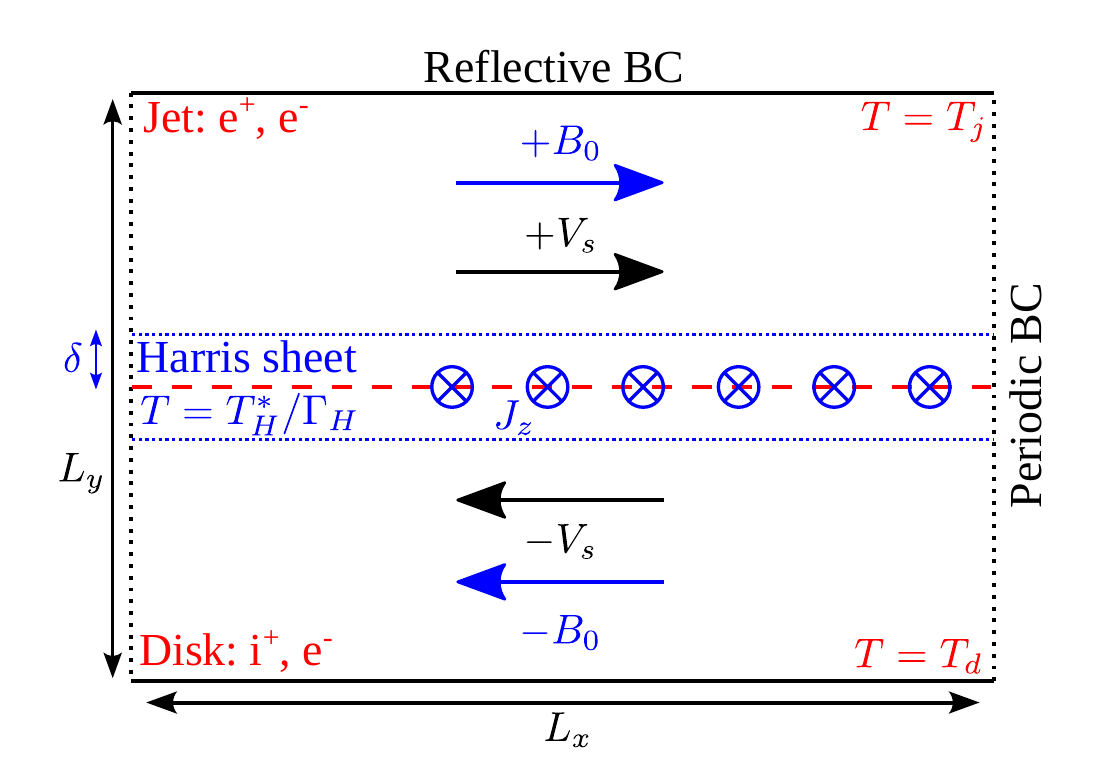}
    \caption{Sketch of the simulation setup. The red dashed line separates the pair plasma on the jet side (upper half) from the electron-ion plasma on the disk side (bottom half). Shear flow is parallel to the magnetic field on each side. ``BC'' stands for ``boundary conditions''.}
    \label{fig::fig1}
\end{figure}

In this work, we employ {\tt Zeltron}, an electromagnetic PIC code developed by \citet{Cerutti2013, 2019ascl.soft11012C}. Our aim is to simulate a current layer that would have been locally formed by instabilities on a global jet-disk interface using a 2D local Cartesian box.  In our simulations, the initial plasma is composed of pairs above the reconnection layer, representing the ``jet side'', while the plasma below is composed of electrons and ions, representing the ``disk side''. As shown by GRMHD simulations (e.g., \citealt{Vos2023}), the magnetic field strength is expected to be similar on each side of the interface. The magnetization contrast should then be only imposed by the composition and/or density asymmetry. We therefore impose the same initial magnetic field strength on both sides of the reconnection layer. This is another difference (in addition to the plasma composition and shear) from the setup used by \citet{Mbarek2022}. 

Our boundary conditions are reflective on the upper and lower sides of the box and periodic on the right and left sides. These conditions apply for both the fields and the particles. For the initial conditions, we adapt the classic relativistic Harris setup with no guide (out-of-plane) field component, which involves an initial antiparallel magnetic field supported by an equatorial current layer (e.g., \citealt{Kirk2003}). This initial current sheet is razor thin (on the plasma skin depth scale, defined below): perhaps much thinner than in a real jet-disk interface. This merely helps kickstart reconnection, allowing us to study the nonlinear evolution of the current sheet in its more developed and effectively thicker stage.

The initial Harris magnetic field profile is given by
\begin{equation}
    \textbf{B}(y)=B_0 \tanh \left( \frac{y-L_y/2}{\delta} \right) \textbf{e}_x,
    \label{eq::Binit}
\end{equation}
where $\delta$ is the initial current layer thickness, $L_y$ ($L_x$) is the height (width) of the box, and $\textbf{e}_x$ is the unit vector along the $x$-direction. This magnetic field is supported by a layer of positively and negatively charged particles counterstreaming, or \textit{drifting}, in the out-of-plane (i.e.,~$z$-) direction. Their bulk drift velocity is set to $V_H/c = 0.6$, giving the current density,
\begin{equation}
    \textbf{J}=-2 e V_H n_H(y) \textbf{e}_z,
\end{equation}
where $e$ is the positron charge. The drifting particle number density reads
\begin{equation}
    n_{H}(y) = n_{H, 0} \cosh^{-2}\left( \frac{y-L_y/2}{\delta} \right),
\end{equation}
where $n_{H, 0}=c B_0/8\pi e V_H \delta$. In symmetric reconnection studies, the box is filled with a uniform background plasma. However, in our scenario, the characteristics of the background plasma vary depending on the vertical position in the box. In the upper half of the box ($y>L_{y}/2$, i.e., the jet side), the background plasma is composed exclusively of pairs of total mass density $2m_e n_j$, while in the bottom half ($y<L_{y}/2$, i.e., the disk side), it is composed of electrons and ions of total mass density $(m_e+m_i)n_d=(1+\mu)m_e n_d$, where $\mu=m_{i}/m_e$ is the ion-to-electron mass ratio. On the jet (disk) side, the plasma moves along field lines with bulk velocity $\textbf{V}_s= V_s\textbf{e}_x$ ($-V_s\textbf{e}_x$) with a corresponding Lorentz factor~$\Gamma_s = (1 - V_s^2/c^2)^{-1/2}$: a symmetric shear with an abrupt transition at the midplane.

In order for the simulation to start in equilibrium, we need to ensure pressure balance both between the current sheet and the upstream magnetic field, and between the jet and the disk. This requires
\begin{align}
   \frac{B^2_0}{8\pi} &= 2 n_{H, 0} \frac{k_B T^*_H}{\Gamma_H} \quad \mathrm{and} \\
    n_{j} k_B T_j &= n_{d} k_B T_d \, ,
\end{align}
where $T_j$ and $T_d$ represent the respective initial temperatures of the jet and disk plasmas in the laboratory frame, $k_B$ is the Boltzmann constant and $T^*_{H}$ is the drifting particles' comoving temperature (considering a Lorentz factor $\Gamma_H=(1-V^2_H/c^2)^{-1/2}$). We note that the plasma-$\beta$ parameter, defined as $\beta = 16 \pi n k_B T / B^2_0$, has an identical value for both the jet and disk sides (see \cite{Mbarek2022} for a $\beta$ asymmetric setup), but the magnetization in the laboratory frame, defined as $\sigma_j = B^2_0/8\pi \Gamma_s n_j m_e c^2$ for the jet and $\sigma_d = B^2_0/4\pi \Gamma_s n_d (m_e + m_i) c^2$ for the disk, changes because of the composition contrast. A global sketch of the initial setup is provided in Fig.~\ref{fig::fig1}.

We need the initial particle energy distribution functions in order to impose the previously defined macroscopic quantities. They follow relativistic drifting Maxwellians as defined by \citet{Swisdak2013}. Therefore, we have to initialize two distribution functions for the current layer (pairs and ions), two for the disk side (electrons and ions) and one for the jet side (pairs). We define four species of particles: the positrons (in the jet region), the ions (of charge~$+e$ in the disk region) and the electrons in each region. Their respective density profiles $n^{e,p,i}_{d,j}$ are
\begin{align}
    n^{e}_j(y) &= n_j H(y-L_y/2)+ f n_H(y), \notag \\
    n^{e}_d(y) &= n_d H(L_y/2-y)+ (1-f) n_H(y), \notag\\
    n^{p}_j(y) &= n_j H(y-L_y/2)+ f n_H(y), \notag \\
    n^{i}_d(y) &= n_d H(L_y/2-y)+ (1-f) n_H(y), \notag
\end{align}
where $H$ is the Heaviside step function, and the parameter $f$ represents the mixing fraction in the Harris sheet. For $f=0$ ($f=1$), the initial current layer is exclusively composed of disk (jet) plasma. This parameter only sets the initial state of our setup, and we have checked that it does not affect the simulation behavior. Thus, we set its value to $f=1/2$. When defining the particles' momenta, we need to decide whether we consider them to be part of the Harris sheet, in which case we assign them a drift velocity $\pm V_{H}\textbf{e}_z$, or in the disk/jet, in which case we assign them a shear bulk motion $\pm V_s\textbf{e}_x$. For a particle at a given coordinate~$y$, the following number is computed:
\begin{equation}
    P(y)=\frac{n_{d,j} H(\pm y \mp L_y/2)}{n^{e,p,i}_{d,j}(y)}.
\end{equation}
Subsequently, a random number between 0 and 1 is drawn. If it exceeds $P(y)$, then the particle is considered to follow the Harris layer energy distribution function; otherwise, it will have a background energy distribution function. It can be easily verified that, far from the current sheet, $P(y) \to 1$ so that the particle will almost always follow a background distribution. Conversely, in the current sheet it will likely follow a Harris distribution.

Following \citet{Werner2018}, a small (one percent) tearing-like perturbation is applied to the initial magnetic field (Eq.~\ref{eq::Binit}). This procedure serves two purposes: (i) it speeds up the onset of reconnection, and (ii) it predefines the location of the main X- and O-points during the active phase of the simulation and facilitates the measurement of the reconnection rate as shown in Sect.~\ref{sec:rec_meas} hereafter. With this perturbation, the initial out-of-plane component of the magnetic vector potential reads
\begin{align}
\begin{split}
    A_z = & B_0 \delta \left\{ \ln\left[ \cosh \left( \frac{y-L_y/2}{\delta} \right) \right]   - \ln\left[ \cosh \left( \frac{L_y}{2\delta} \right) \right] \right\} \\ 
    &\times \left\{ 1 + 0.01 \cos \left( 2\pi \frac{x-3 L_x/4}{L_x} \right) \cos^2 \left( \pi \frac{y-L_y/2}{L_y} \right) \right\}.
\end{split}
\end{align}
The dimensions of the cells in each direction, $\Delta x$ and $\Delta y$, are set to be equal and the time step, $\Delta t$, is set at $0.99$ times the critical Courant-Friedrichs-Lewy time step. We initialize the simulations with 16 particles per species per cell.

In summary, the setup allows for the manipulation of several free parameters, including plasma $\beta$, the disk magnetization $\sigma_d$, the jet-disk density ratio $n_j/n_d$, and the shear flow velocity $V_s$. For the remainder of this study, we set $n_j = n_d$ for numerical convenience, thus only imposing a magnetization asymmetry through the mass ratio between pairs and ions. This assumption may not be very realistic in the context of a jet-disk boundary layer where a strong number density contrast is expected, but it still captures a mass density contrast, $2 n_j m_e\ll n_d (m_i+m_e)$.

\subsection{Sets of simulations}

\begin{table*}[h]
    \centering
    \caption{Full list of simulations performed in this study. Top: Asymmetric setup exploring the $(\beta,\sigma_d)$ parameter space. Middle: Corresponding symmetric electron-ion reconnection simulations included for comparison with the asymmetric runs. Bottom: Shear simulation parameters.}
    \setlength\tabcolsep{0pt}
    \begin{tabular*}{\linewidth}{@{\extracolsep{\fill}} ccccccccc }
    \hline\hline
        Simulation name & $L_x/\rho_0$ & $L_y/\rho_0$ & Grid cells & $\mu$ & $\sigma_d$ & $\beta$ & $U_s/\sqrt{\sigma_d}$  & Measured $\langle \beta_{\rm rec} \rangle$  \\
        \hline 
         \texttt{SD01B0001}& 6144 & 3072 & $16384 \times 8192$ & 100 & $10^{-1}$ & $10^{-3}$ & 0 & $2.9 \times 10^{-2}$ \\  
         \texttt{SD01B001}& 6144 & 3072 & $16384 \times 8192$ & 100 & $10^{-1}$ & $10^{-2}$ & 0 & $3.1 \times 10^{-2}$ \\
         \texttt{SD01B01}& 6144 & 3072 & $16384 \times 8192$ & 100 & $10^{-1}$ & $10^{-1}$ & 0 & $2.2 \times 10^{-2}$ \\
         \texttt{SD03B0001}& 6144 & 3072 & $16384 \times 8192$ & 100 & $3\times 10^{-1}$ & $10^{-3}$ & 0 & $5.4 \times 10^{-2}$\\  
         \texttt{SD03B001}& 6144 & 3072 & $16384 \times 8192$ & 100 & $3\times 10^{-1}$ & $10^{-2}$ & 0 & $5.5 \times 10^{-2}$ \\
         \texttt{SD03B01}& 6144 & 3072 & $16384 \times 8192$ & 100 & $3\times 10^{-1}$ & $10^{-1}$ & 0 & $5.0 \times 10^{-2}$\\
         \texttt{SD1B0001}& 6144 & 3072 & $16384 \times 8192$ & 100 & 1 & $10^{-3}$ & 0 & $9.7 \times 10^{-2}$ \\  
         \texttt{SD1B001}& 6144 & 3072 & $16384 \times 8192$ & 100 & 1 & $10^{-2}$ & 0 & $7.7 \times 10^{-2}$ \\
         \texttt{SD1B01}& 6144 & 3072 & $16384 \times 8192$ & 100 & 1 & $10^{-1}$ & 0 & $7.2 \times 10^{-2}$ \\
         \hline
         \texttt{S01B0001SYM}& 6144 & 3072 & $16384 \times 8192$ & 100 & $10^{-1}$ & $10^{-3}$ & 0 & $2.2 \times 10^{-2}$\\ 
         \texttt{S1B0001SYM}& 6144 & 3072 & $16384 \times 8192$ & 100 & 1 & $10^{-3}$ & 0 & $7.8 \times 10^{-2}$ \\
         \hline
         \texttt{SD1US0}& 512 & 256 & $8192 \times 4096$ & 1 & $1$ & $5\times 10^{-2}$ & 0 & $4.1 \times 10^{-2}$ \\ 
         \texttt{SD1US1}& 538 & 269 & $8192 \times 4096$ & 1 & $1$ & $5\times 10^{-2}$ & 1/3 & $5.2 \times 10^{-2}$\\ 
         \texttt{SD1US2}& 615 & 308 & $8192 \times 4096$ & 1 & $1$ & $5\times 10^{-2}$ & 2/3 & $3.5 \times 10^{-2}$\\ 
         \texttt{SD1US3}& 724 & 362 & $8192 \times 4096$ & 1 & $1$ & $5\times 10^{-2}$ & 1 & $2.6 \times 10^{-2}$\\ 
         \texttt{SD1US4}& 853 & 427 & $8192 \times 4096$ & 1 & $1$ & $5\times 10^{-2}$ & 4/3 & $1.9 \times 10^{-2}$\\ 
         \texttt{SD1US6}& 1145 & 572 & $8192 \times 4096$ & 1 & $1$ & $5\times 10^{-2}$ & 2 & $1.4 \times 10^{-2}$\\ 
         \texttt{SD10US0}& 4096 & 2048 & $8192 \times 4096$ & 1 & $10$ & $5\times 10^{-2}$ & 0 & $8.9 \times 10^{-2}$\\ 
         \texttt{SD10US1}& 5951 & 2976 & $8192 \times 4096$ & 1 & $10$ & $5\times 10^{-2}$ & 1/3 & $7.8 \times 10^{-2}$\\ 
         \texttt{SD10US2}& 9957 & 4779 & $8192 \times 4096$ & 1 & $10$ & $5\times 10^{-2}$ & 2/3 & $4.4 \times 10^{-2}$\\ 
         \texttt{SD10US3}& 13585 & 6792 & $8192 \times 4096$ & 1 & $10$ & $5\times 10^{-2}$ & 1 & $5.0 \times 10^{-2}$\\ 
         \texttt{SD10US4}& 17749 & 8875 & $8192 \times 4096$ & 1 & $10$ & $5\times 10^{-2}$ & 4/3 & $3.4 \times 10^{-2}$\\ 
         \texttt{SD10US6}& 26227 & 13114 & $8192 \times 4096$ & 1 & $10$ & $5\times 10^{-2}$ & 2 & $1.9 \times 10^{-2}$\\ 
         \hline
    \end{tabular*}
    \label{tab:table1}
\end{table*}

Given our setup, this work consists of two series of simulations. In the first, we study the effect of composition asymmetry on magnetic reconnection without velocity shear (i.e., setting $V_s = 0$). This asymmetry leads to a magnetization contrast since, for~$\mu \gg 1$, the jet magnetization is $\sigma_j \simeq \mu \sigma_d /2$. We set $\mu=100$ so that the magnetization ratio is $\sigma_j/\sigma_d \simeq 50$, and we explore the parameter space $(\beta,\sigma_d)$ in the 9 simulations listed in Table \ref{tab:table1}. We adopt plasma conditions previously studied by \citet{Ball2018} and similar to those observed at jet-disk interfaces in GRMHD simulations, where the ion magnetization is transrelativistic ($\sigma_d \lesssim 1 $) while that for pairs is ultrarelativistic ($\sigma_j \gg 1$). In addition to our 9 asymmetric simulations, we present 2 symmetric electron-ion simulations for comparison where the jet's plasma has the same properties as the disk's. The resolution is fixed in terms of the nominal electron Larmor radius $\rho_0 = m_e c^2 /e B_0 = 8\Delta x /3$. Thus, the skin-depth ranges from $d_e = c/\omega_{pe} \sim 8 \Delta x$ to $ d_e \sim 27 \Delta x$, where $\omega_{pe}=\sqrt{4 \pi n_d e^2/m_e}$ is the electron plasma frequency. The initial current sheet width is set to $\delta \simeq 5 \Delta x$.

The second set of simulations focuses on the effect of a velocity shear along the reconnection layer with a symmetric pair plasma ($\mu = 1$). Here, we fix $\beta = 5\times 10^{-2}$ and the magnetization,~$\sigma_d$, is set to either $1$ or $10$. For both values of $\sigma_d$, we scan the dimensionless shear four-velocity, $U_s = \Gamma_s V_s/c $, from 0 to $2U_A$, where $U_A=\sqrt{\sigma_d}$ is the relativistic Alfv\'en four-velocity. The resolution is set such that $\Gamma_s \rho_0 = 16 \Delta x$ for $\sigma_d = 1$, and $\Gamma_s \rho_0 = 2 \Delta x$ for $\sigma_d = 10$. This ensures that the hot skin depth $d_{e,h}=\sqrt{\Gamma_s}d_e$ is conserved for all simulations sharing the same $\sigma_d$ value: $d_{e,h} \sim 6 \Delta x$ ($\sigma_d=10$), $d_{e,h} \sim 16 \Delta x$ ($\sigma_d=1$). The current sheet width is set to $\delta = 4\Delta x$. A summary of the simulation parameters is provided in Table \ref{tab:table1}.

\subsection{Measurement of the reconnection rate}
\label{sec:rec_meas}

We measure the reconnection rate based on the evolution of the reconnected magnetic flux, denoted as $\Phi_{\rm rec}$, and defined by the equation,
\begin{equation}
    \Phi_{\rm rec} = A_z(X)-A_z(O).
\end{equation}
Here, ``X'' represents the location of the major X-point, which corresponds to the maximum of $A_z$ in the midplane. On the other hand, ``O'' denotes the location of the major O-point at the center of the largest plasmoid; it coincides with the minimum of $A_z$ in the midplane. Using Faraday's law, the instantaneous dimensionless reconnection rate, denoted as $\beta_{\rm rec}$, can then be expressed as,
\begin{equation}
    \beta_{\rm rec} = \frac{1}{c B_0} \frac{\textrm{d}\Phi_{\rm rec}}{\textrm{d}t}.
\end{equation}
Following the methodology outlined by \citet{Werner2018}, we calculate a mean reconnection rate, denoted as $\langle \beta_{\rm rec} \rangle$, by averaging the previously defined instantaneous rate over the time interval between $t_{\rm 20\%}$ and $t_{\rm 40\%}$. Here, $t_{\rm 20\%}$ and $t_{\rm 40\%}$ correspond to the instants when $20\%$ and $40\%$ of the initial magnetic flux in half of the box, $\Phi_0 = B_0 L_y/2$, has reconnected. By measuring in this time interval, we aim to isolate the intrinsic reconnection rate, avoiding artificial contributions from both the initial phase, when reconnection is still ramping up, and from the late-time phase, when reconnection starts to slow down due to the finite flux available in the box. Because of the intermittent nature of magnetic reconnection, there is a scatter of about $10\%$ to $15\%$ in our measurements of the reconnection rate.

\section{Composition asymmetry results}\label{sec:SpecAsym}

\subsection{Kinematics and global behavior}\label{sec:SpecAsym_global}

\begin{figure*}[h]
    \centering
    \includegraphics[width=\textwidth]{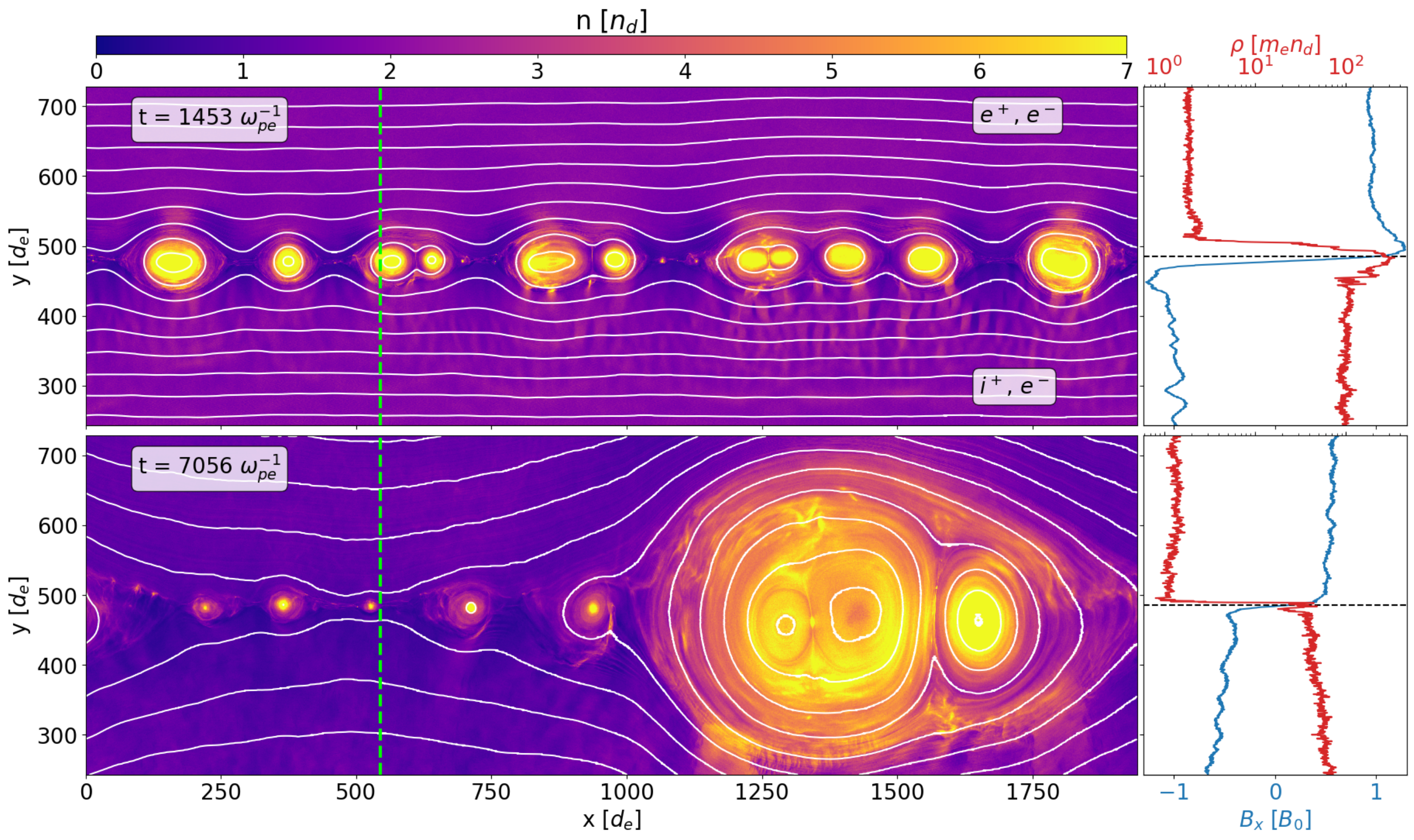}
    \caption{Plasma number density in the composition asymmetric setup (no shear) shown in the middle of the reconnection process (top left panel, $\Phi_{\rm rec} \sim 0.15 \Phi_0 $) and towards the saturation stage (bottom left panel, $\Phi_{\rm rec} \sim 0.45 \Phi_0 $) for $\sigma_d = 0.1$, $ \beta = 10^{-3}$ and $\sigma_j = 5$ (run {\tt SD01B0001}). White contours represent magnetic field lines. We provide on the top and bottom right panels a plot of different quantities for a slice shown by the green dashed line on the density maps. The blue curve represents the x-component of the magnetic field and the red curve represents the plasma mass density $\rho$ (in logarithmic scale). The box midplane is represented by the black dashed lines. For the time evolution, see the animation on the journal website.} 
    \label{fig::fig2}
\end{figure*}

We show two representative snapshots of the plasma density and field lines from our composition-asymmetric runs in Fig.~\ref{fig::fig2}. At first glance, there are no noticeable differences with previous symmetric studies: the current sheet becomes quickly unstable to the tearing mode, creating a chain of merging plasmoids separated by X-points. However, a closer look at the system -- and in particular its time evolution -- reveals that fast magnetosonic waves launched during plasmoid mergers propagate faster and on shorter lengthscales on the pair side than on the electron-ion side, an effect resulting from the higher~$\sigma$ of the pair plasma. This wave-speed asymmetry leads to differing plasma response times on each side, which provokes an oscillatory vertical motion of the current layer that is absent in symmetric simulations (see the full time evolution of the simulation in the animation provided on the journal website).

As opposed to previous studies of both relativistic and nonrelativistic asymmetric magnetic reconnection \citep{Murphy2012,Mbarek2022}, the shape of the plasmoids seems rather symmetric. This behavior can be explained by the fact that the magnetic field strength is symmetric in our setup. The magnetic pressure is thus the same on both sides of the reconnection layer, resulting in the absence of strong plasmoid deformation.

We measure reconnection rates in our asymmetric configurations that are slightly higher ($\sim 30$\% more) than in our corresponding symmetric electron-ion plasma simulations. For instance, we obtain $\langle \beta_{\rm rec} \rangle \simeq 0.029$ for the asymmetric run with $\sigma_d = 0.1$ and $\beta = 10^{-3}$ while the corresponding symmetric simulation showed $\langle \beta_{\rm rec} \rangle \simeq 0.022$ (see Table \ref{tab:table1}). Even so, the rate stays much smaller than what would be expected for symmetric pair-plasma reconnection (i.e., $\beta_{\rm rec} \sim 0.1$). This suggests that the side with the lowest magnetization primarily dictates the overall reconnection rate, as previously demonstrated in the context of density and magnetic field asymmetries by \citet{Mbarek2022}.

\subsection{Energy partition and mixing}
\label{sec:energypartition}

\begin{figure*}
    \centering
    \includegraphics[width=\textwidth]{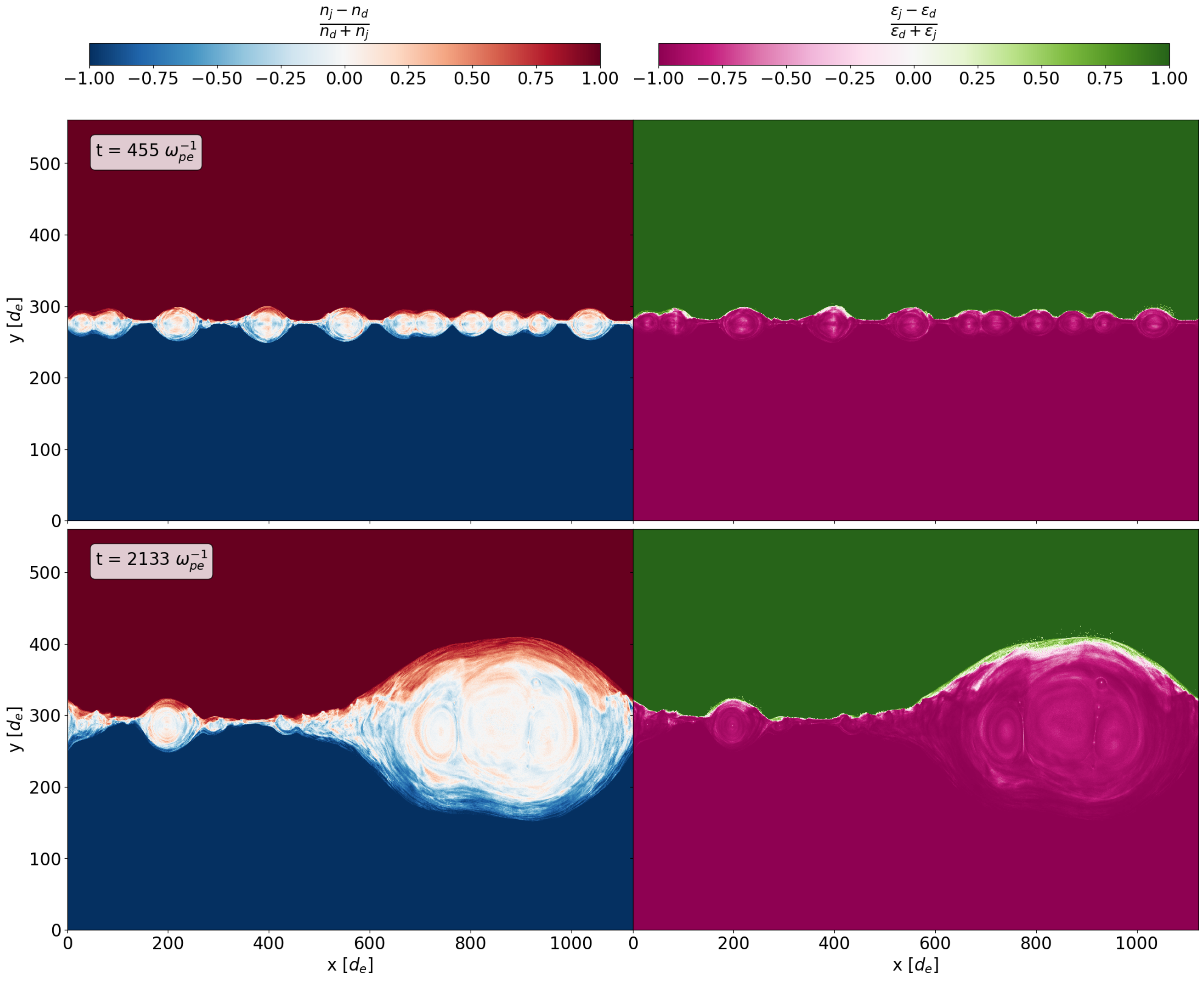}
    \caption{Plasma number density mixing (left panels) defined as $(n_j-n_d)/(n_d+n_j)$, and plasma kinetic energy density mixing (right panels) defined as $(\epsilon_j-\epsilon_d)/(\epsilon_d+\epsilon_j)$, where $\epsilon_{j/d}=\gamma n_{j/d} m c^2$, in the early (top panels, $\Phi_{\rm rec} \sim 0.15 \Phi_0$) and late (bottom panels, $\Phi_{\rm rec} \sim 0.5 \Phi_0$) stages of the simulation. Here, $\sigma_d=0.3$ and $\beta=10^{-2}$ (run {\tt SD03B001}).}
    \label{fig:SpecAsym_mix}
\end{figure*}

We investigate the particle and energy mixing between the disk and the jet shown in the left panels of Fig.~\ref{fig:SpecAsym_mix}. The density mixing between the jet and the disk is very efficient at the X-points and in the plasmoids. This behavior may explain the previously observed dynamics of the reconnection process: because reconnection efficiently mixes particles, the magnetization in the reconnection layer is locally the one imposed by the disk ions, since, after mixing, these particles control the local mass density. Looking at the energy mixing in Fig.~\ref{fig:SpecAsym_mix} (right panels), we observe that the energy density of the disk significantly prevails within the plasmoids. This dominance of the disk energy, coupled with the homogeneous mixing of particles, implies that the disk undergoes more substantial energization compared to the jet plasma during the reconnection process. Additionally, we remark that a few highly energetic protons manage to get to the upper edge of plasmoids. We speculate that, in 3D, where plasmoids have been shown to be less efficient at confining high-energy particles \citep{Zhang2021, 2023ApJ...959..122C}, this tendency would allow particles to escape plasmoids even more effectively and contaminate the jet.

\begin{figure}
    \centering
    \includegraphics[width=\columnwidth]{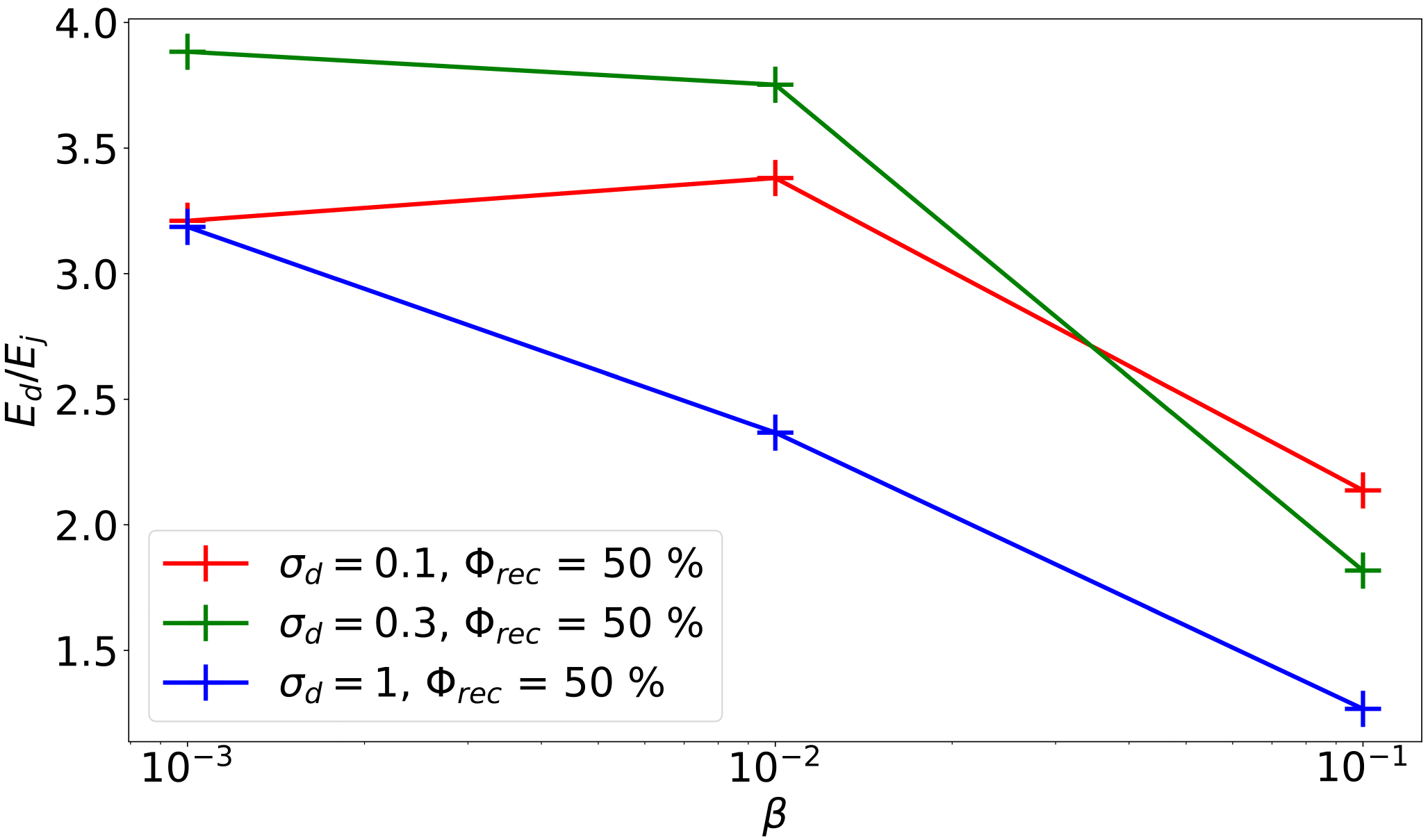}
    \caption{Energy partition between the electron-ion (disk) plasma and the pair (jet) plasma at the stage when half of the initial magnetic flux has reconnected, $\Phi_{\rm rec}=0.5\Phi_0$.}
    \label{fig:SpecAsym_energypartition}
\end{figure}

We plot the ratio of the jet-to-disk particles' total kinetic energies in Fig.~\ref{fig:SpecAsym_energypartition} for our entire composition-asymmetric simulation campaign.
The total kinetic energies for disk and jet particles are defined as
\begin{align}
E_d &= \int (\gamma-1) \left(\frac{\mathrm{d}N^d_e}{\mathrm{d}\gamma} + \mu \frac{\mathrm{d}N^d_i}{\mathrm{d}\gamma} \right) \mathrm{d} \gamma \quad \textrm{and} \\
E_j &= \int  (\gamma-1) \left(\frac{\mathrm{d}N^j_e}{\mathrm{d}\gamma} + \frac{\mathrm{d}N^j_p}{\mathrm{d}\gamma} \right) \mathrm{d} \gamma.
\end{align}

In order to compare simulations with different reconnection rates -- and thus varying time evolution -- we compute this ratio once 50\% of the initial magnetic flux has reconnected (the results are unchanged if we consider the instants when 30\% and 40\% of the initial flux is reconnected). The first conclusion is that in every case, the disk retrieves more energy than the jet from the reconnection event: up to four times more for the lowest plasma-$\beta$ simulations, $\beta=10^{-3}$. Second, higher magnetization ($\sigma_d = 1$) means a more symmetric behavior between the pairs and ions, and thus explains why the energy ratio decreases when $\sigma_d = 1$. With a higher value of $\beta$, the plasma's initial thermal energy is more significant compared with the magnetic energy. Hence the energization during magnetic reconnection is less significant and the energy ratio is thus less affected. However, the fact that $\sigma_i = 0.3$ seems to create slightly more asymmetry than $\sigma_i=0.1$ at low-$\beta$ remains unexplained.

We emphasize the fact that most of the asymmetry in~$E_d/E_j$ is due to the ions acquiring much more energy from the reconnection process, as previously observed by \citet{Werner2018, Ball2018} in electron-ion reconnection studies. Despite this qualitative similarity with the symmetric case, the preferential energization of ions over electrons is quantitatively more extreme in the presence of asymmetry with respect to findings of earlier symmetric electron-ion reconnection work. We measure, for instance, in our $\sigma = 0.1$ and $\beta=10^{-3}$ simulation, that the ion kinetic energy ends up $\sim 5.6$ times higher than that of the disk electrons (i.e., a higher ratio than $E_d/E_j$ in Fig. 4 because here we compare the disk ions to only the disk electrons). In contrast, for our corresponding symmetric simulation, we measure this ratio as 3.4, consistent with \cite{Werner2018}. This extra preferential ion energization tends to vanish for higher magnetizations.

\subsection{Spectra}

\begin{figure}
    \centering
    \includegraphics[width=\columnwidth]{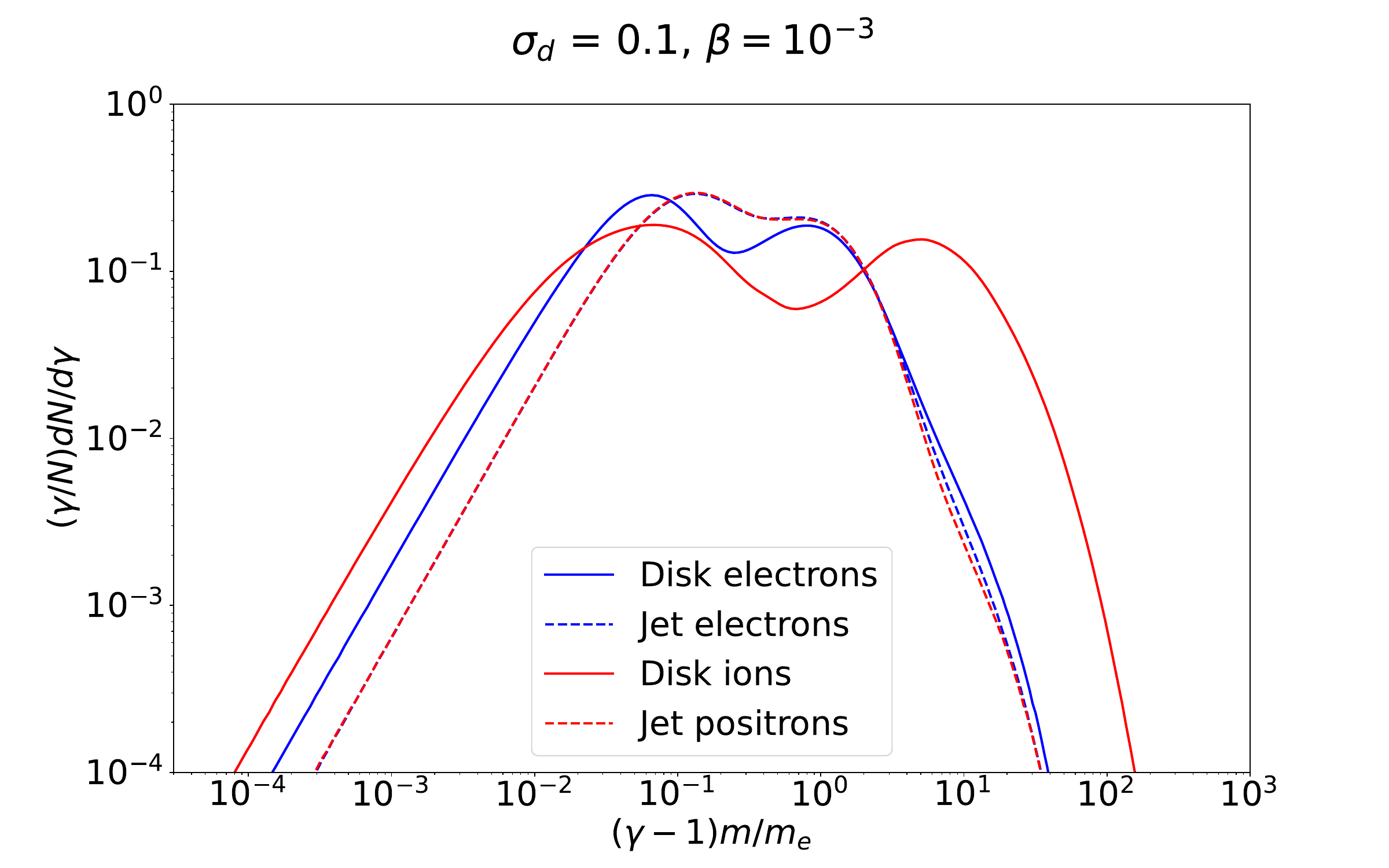}
    \includegraphics[width=\columnwidth]{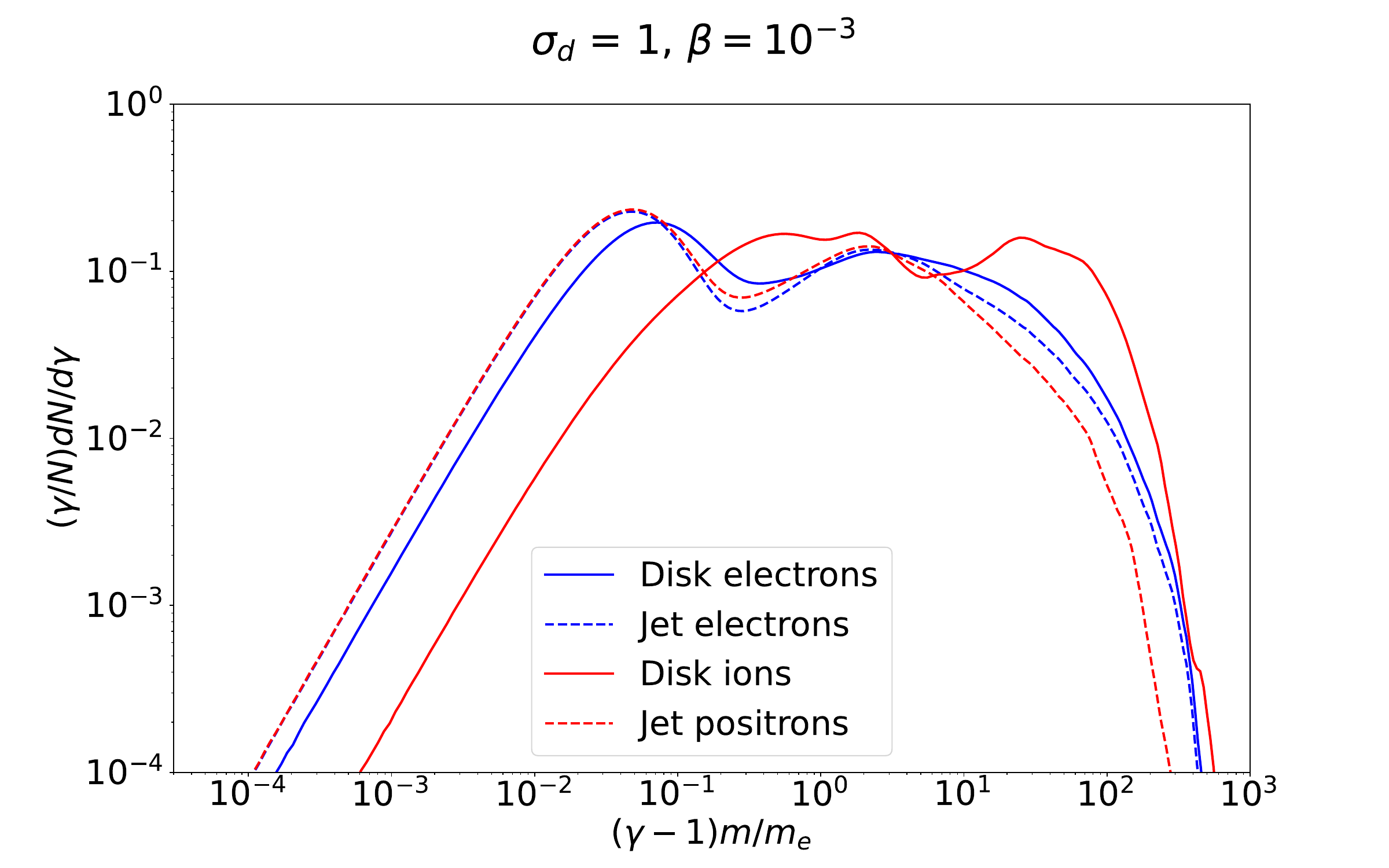}
    \caption{Particle energy distributions for all species at the end of the $\sigma_d = 0.1$ ({\tt SD01B0001}, top panel) and $\sigma_d = 1$ ({\tt SD1B0001}, bottom panel) runs, for both $\beta = 10^{-3}$.}
    \label{fig:SpecAsym_spectra}
\end{figure}

We further investigate the asymmetry in plasma energy gain during the reconnection process by looking at the particle energy distributions (Fig.~\ref{fig:SpecAsym_spectra}). Looking first at the $\sigma_i = 0.1$, $\beta=10^{-3}$ run, we notice that the ions reach higher energies than the leptons ($\sim 3$ times more). Given that the magnetization is less than one, a distinct nonthermal tail is not apparent. These two observations are in accord with earlier symmetric studies \citep{Melzani2014b, Werner2018, Ball2018}. Instead of a non thermal tail, the particle spectra are composed of two nearly thermal humps: a low-energy hump representing the initial thermal distribution, and a high-energy component corresponding to the plasma energized by reconnection. The gap of a factor $\sim10$ between the high-energy electron and ion peaks is consistent with the lower efficiency for the lepton heating observed in Sec.~\ref{sec:energypartition}.  If the leptons were instead heated as much as the ions, both species should reach a final energy of $(\langle \gamma \rangle - 1) m/m_e \sim \sigma_i m_i/m_e \sim 10$. However, this is only consistent with the energy of the ion peak ($\sim 10$); the lower electron peak ($\sim 1$) can only be explained by a reduced lepton heating efficiency with respect to ions. 
Examining next the $\sigma_d = 1$ regime, nonthermal tails with slightly differing slopes start appearing for the leptons. The cutoff energy seems to be similar for all species (electrons and ions), however the fraction of ions getting to higher energies is significantly higher.

\begin{figure}
    \centering
    \includegraphics[width=\columnwidth]{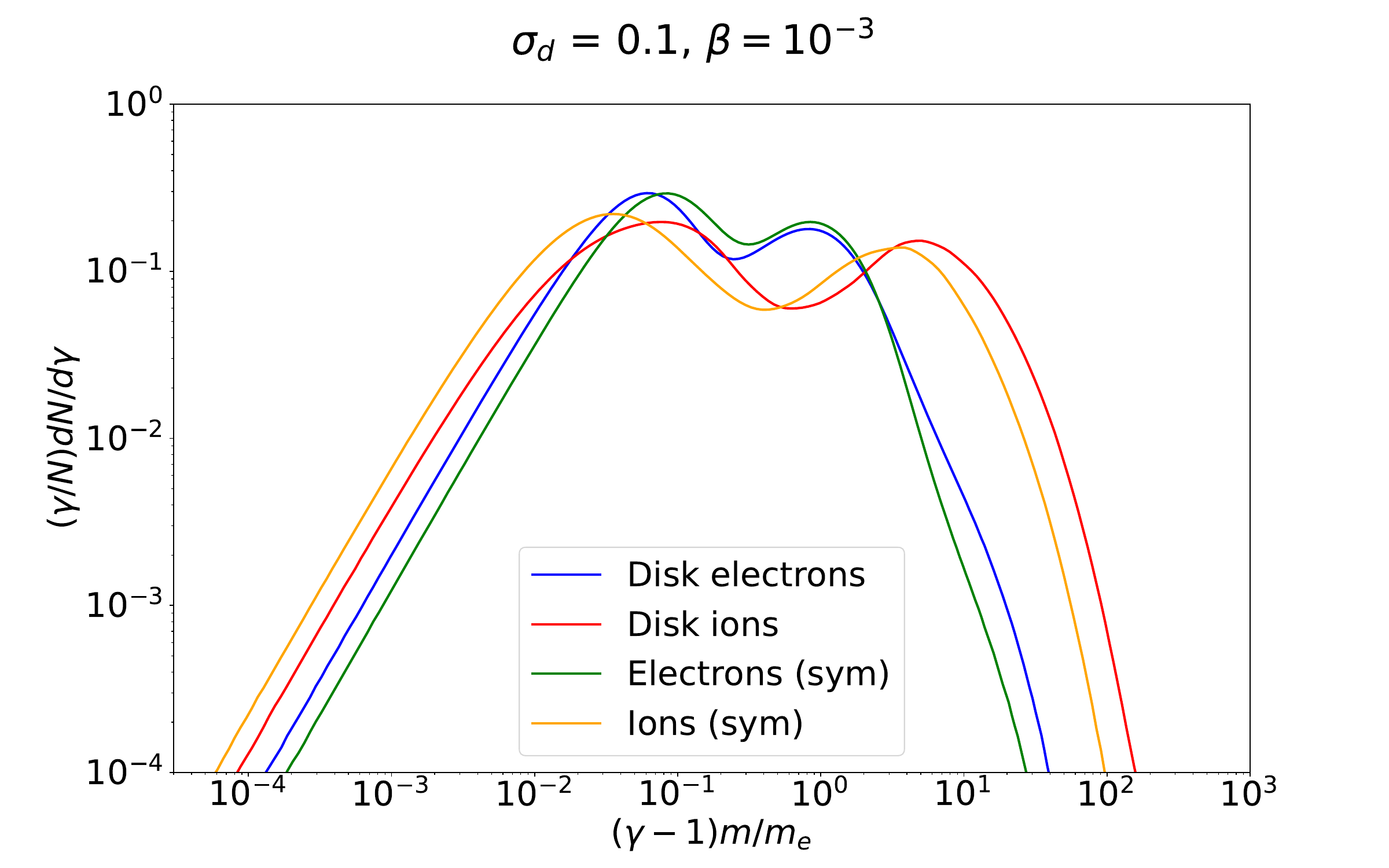}
    \caption{Same as the top panel in Fig.~\ref{fig:SpecAsym_spectra} but with the plasma properties identical on both sides (i.e., symmetric reconnection with $\sigma_j = \sigma_d = 0.1$ and $\beta=10^{-3}$, run {\tt S01B0001SYM}).}
    \label{fig:SpecAsym_spectra_sym_asym}
\end{figure}

We compare the asymmetric and symmetric runs with~$\sigma_d=0.1$ and~$\beta=10^{-3}$ in Fig.~\ref{fig:SpecAsym_spectra_sym_asym}. Both runs show similar spectral shapes, but the asymmetric particle distribution appears shifted to slightly higher energies. One reason for this shift could be that there are two times more ions for the same amount of magnetic energy in the symmetric simulation compared to the asymmetric run. Overall, despite subtle differences, the asymmetric and corresponding symmetric runs both show a strong bias, as the magnetization decreases, toward ion acceleration at the expense of electrons. The key insight from this analysis is then that it is this bias -- which is evidently independent of the composition asymmetry -- that primarily explains the large energization discrepancy between the jet and disk plasmas.

\section{Sheared reconnection results}\label{sec:Shear}

\subsection{Overall picture}

\begin{figure}
    \centering
    \includegraphics[width=1\columnwidth]{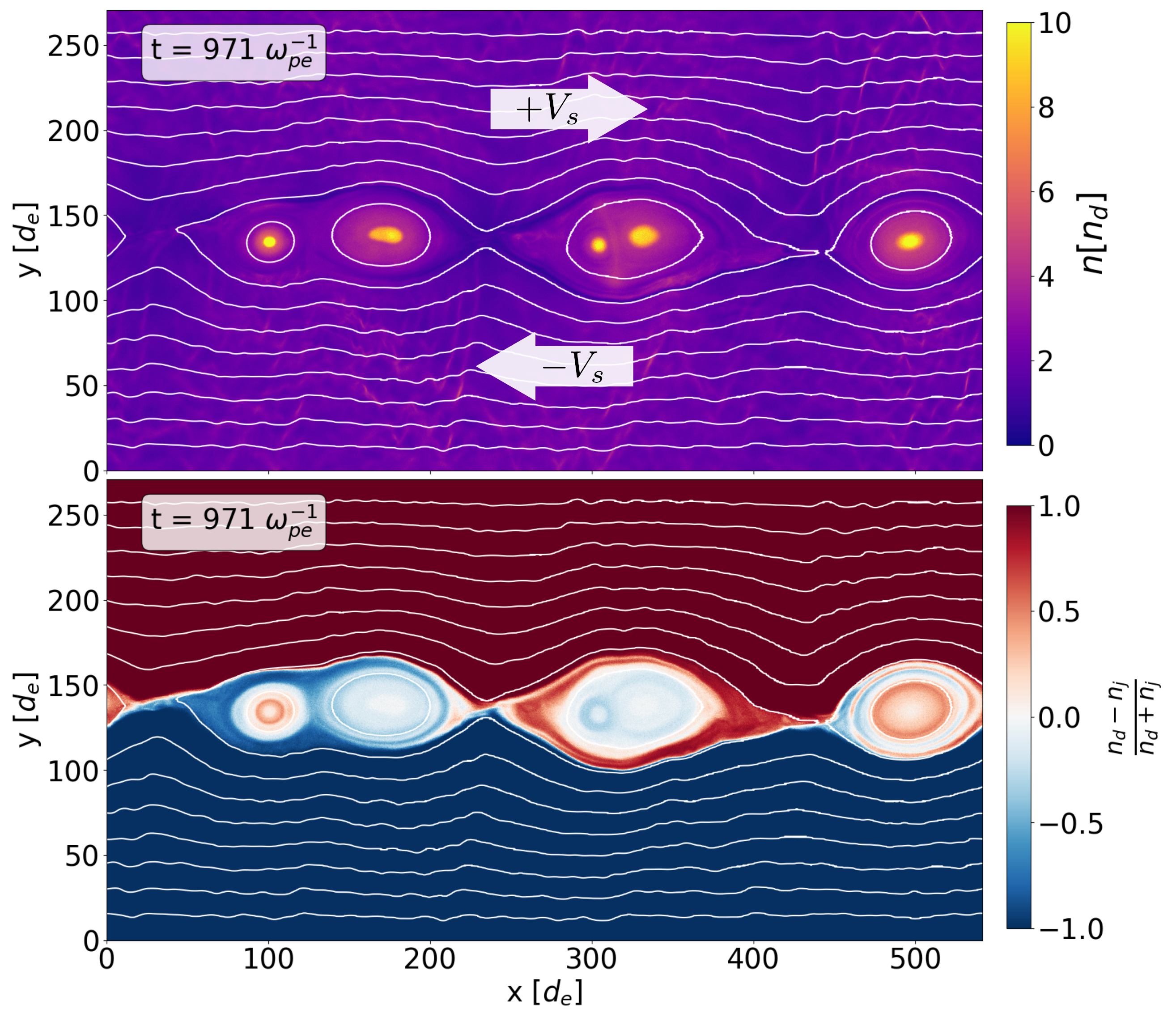}
    \caption{Simulation of pair-plasma relativistic magnetic reconnection with an antisymmetric shear applied along the current layer. The top panel shows the plasma number density; the bottom panel shows the mixing of both upstream plasmas as in Fig.~\ref{fig:SpecAsym_mix}. Here, $\sigma_d = 1$, $U_s = 2 U_A$ (run {\tt SD1US6}) and the white contours represent the magnetic field lines. For the time evolution, see the animation on the journal website.}
    \label{fig:Sheardensmix}
\end{figure}

In our second set of simulations, we return to a a symmetric pair-plasma composition. Here, the current layer is antisymmetrically sheared by the upstream pair plasma, which moves with a bulk velocity~$\pm V_s \textbf{e}_x$. Given that certain simulations involve elevated bulk Lorentz factors ($\Gamma_s \gtrsim 5$), we apply a 9-point digital filter to the current densities at each timestep \citep{Langdon1970} to avoid the numerical Cherenkov effect.

We show a plasma density plot for the case of super-Alfv\'enic shear ($U_s = 2 U_A$) in Fig.~\ref{fig:Sheardensmix}. The propagation of fast magnetosonic waves is affected by the shear: the wave fronts are inclined with respect to the reconnection layer instead of, as occurs when~$V_s = 0$, nearly parallel to it.
Despite the strong bulk motion, X-points and plasmoids are not swept away because the shear is symmetric.

Spiral patterns clearly appear in the density mixing map, as shown on the bottom panel of Fig.~\ref{fig:Sheardensmix} (e.g., in the rightmost plasmoid with center located near~$x = 500 d_e$). In the context of shear, plasma flows into plasmoids from X-points with a preferential velocity direction set by whether the X-point flanks the plasmoid to the right or to the left. Plasma entering from an X-point to the right (left) of a given plasmoid comes preferentially from the leftward-moving (rightward-moving) upstream plasma below (above) the layer. This plasma then follows a freshly reconnected magnetic field line, tracing a spiral trajectory around the core of the growing plasmoid. A plasmoid trapped between two active X-points is thus fed by the upper plasma from its left side and by the lower plasma from its right side, producing, respectively, the red- and blue-colored spirals in Fig.~\ref{fig:Sheardensmix}. Other than such spiral features in the density mixing map, reconnection proceeds in a similar fashion as for an initially static upstream plasma.

According to the linear analysis performed by \citet{Chow2023}, we may expect the appearance of the Kelvin-Helmholtz (KH) instability for simulations where $V_s > V_A$. However, the tearing instability seems to grow more rapidly than the KH instability, and there is no clear sign of KH vortex formation throughout the duration of the simulations. As plasmoids grow during the evolution of the system, the velocity gradient across the midplane decreases, further inhibiting the operation of the KH instability. A sharp velocity gradient is maintained only in the vicinity of X-points. Toward the end of the simulation (e.g. when $\mathrm{d}\Phi_{\rm rec} / \mathrm{d}t \sim 0$), we observe that super-Alfv\'enic shears tend to decelerate to sub-Alfv\'enic speeds, due in part to the growth of large plasmoids, the periodic boundary conditions, and the finite box size. We believe that an outflowing box setup would mitigate this late-time deceleration of the shear, but it would not strongly impact our measurements (presented below) of the reconnection rate and the particle energy distribution. These are performed much earlier in the simulation: before shear-flow deceleration linked to excessively large (periodic-boundary-induced) plasmoids mature.

\subsection{Reconnection rate}

\begin{figure}[h]
    \centering
    \includegraphics[width=\columnwidth]{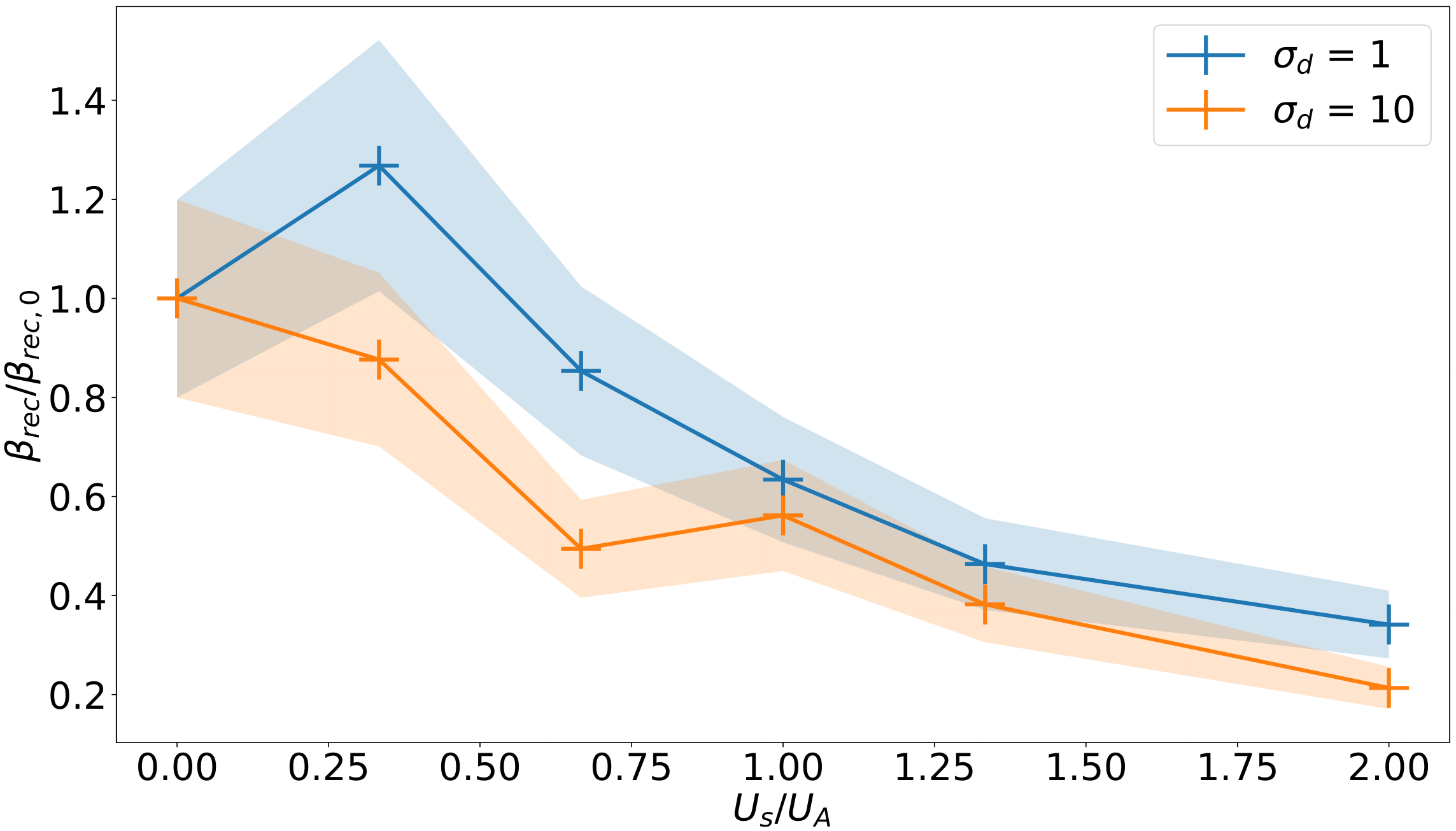}
    \caption{Reconnection rate as a function of the shear-flow 4-velocity, $U_s$, in units of the Alfv\'en 4-velocity,~$U_A=\sqrt{\sigma_d}$. The rates are normalized (separately for~$\sigma_d=1$ and~$\sigma_d=10$) to the $U_s=0$ rate denoted as $\beta_{\mathrm{rec}, 0}$. The error bars represent our estimated relative uncertainty of~$10\%$ through occurrences of a given run as described in Sec.~\ref{sec:rec_meas}.}
    \label{fig:ShearRecRate}
\end{figure}

We compute the reconnection rates for each shear simulation using the method outlined in Sec.~\ref{sec:rec_meas}. Figure~\ref{fig:ShearRecRate} presents the reconnection rate, relative to its $V_s = 0$ reference value, as a function of the shear four-velocity, $U_s = V_s \Gamma_s / c$, normalized to the Alfv\'enic four-velocity, $U_A=\sqrt{\sigma_d}$. Our primary observation is that reconnection slows down with increasing shear speeds, dropping to about 20\% of its reference rate once $U_s$ reaches $2 U_A$. Notably, for super-Alfv\'enic speeds, the reconnection rate appears to follow a similar trend between the $\sigma = 1$ and $\sigma = 10$ cases, suggesting a potential universality in this behavior. We point out, however, a local enhancement of the reconnection rate for modest shear speeds ($U_s=U_A/3$) that occurs only for $\sigma_d=1$. This feature is not clearly understood, but it might not be significant given the~$\sim 10\%$ uncertainty in the rate measurements reported in Sect.~\ref{sec:rec_meas}. This uncertainty is somewhat amplified by the fact that we normalize by the reconnection rate with no shear (which itself is also only accurate to roughly $10\%$). Similar considerations apply to the slight dip observed at $\sigma_d = 10$ ($U_s = 2 U_A /3$). The overall decrease of the reconnection rate with increasing shear-flow velocity may be due to the loss of causal connection between the fast-moving upstream plasma and the nearly static X-points.

The general slowing of reconnection with increasing shear is in qualitative agreement with previous non-relativistic studies \citep{LaBelle-Hamer1995,Cassak2011a,Cassak2011b}. Quantitatively, however, these earlier works consistently indicate a thorough quenching of reconnection, with a very sharp cutoff near the Alfv\'en point. Our relativistic simulations, in contrast, show a much more gradual behavior across this point, and a complete shutdown of magnetic reconnection is not observed. This result is at odds with what has been recently reported by \citet{2024ApJ...964..144P}, where relativistic reconnection is already quenched for shear speeds substantially below the Alfv\'en point. This may be explained by the non-zero guide field or by the small box size used in their simulations.

\subsection{Implications for particle acceleration efficiency}

\begin{figure}
    \centering
    \includegraphics[width=\columnwidth]{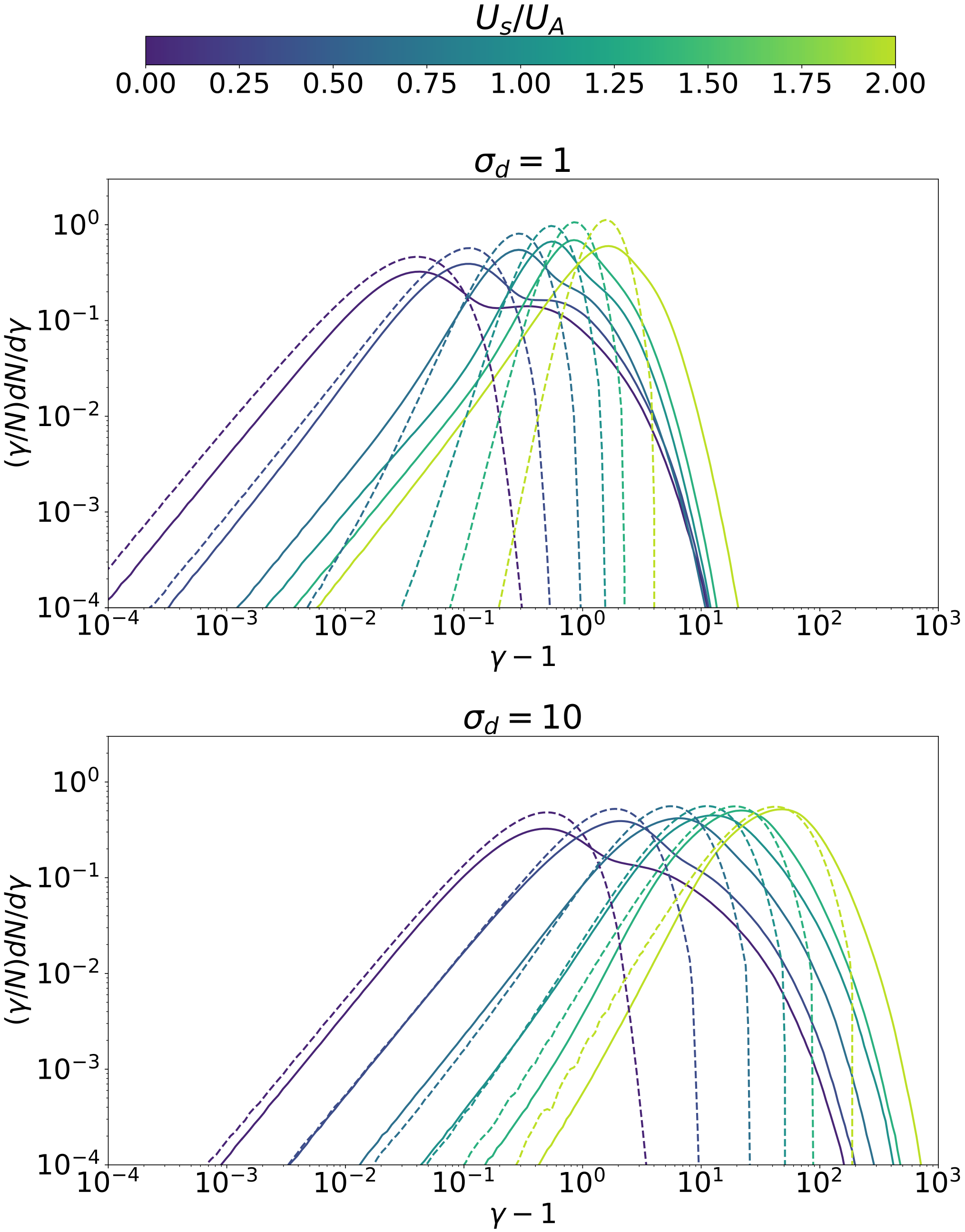}
    \caption{Comparison of the particle energy distributions at a given magnetization for different shear values and $\Phi_{\rm rec}=0.5\Phi_0$. Dashed lines represent the corresponding initial spectra.}
    \label{fig:Shear_spectra}
\end{figure}

To comprehend the impact of shear flow on particle acceleration efficiency, we present, in Fig.~\ref{fig:Shear_spectra}, the particle energy distributions measured when~$\Phi_{\rm rec} =0.5\Phi_0$, which occurs at differing simulation times due to the change in the laboratory-frame reconnection rate reported above. The initial distributions are shown on the figure with dashed lines for comparison. We observe that the high-energy power-law tail steepens with increasing shear. This is certainly true at low shear and seems to hold at high shear as well, though in the latter case~(say, at $U_s \ge 4U_A/3$) the initial plasma energy is so high that an unambiguous nonthermal component is less clear.

We also notice, for $\sigma_d = 1$, an emerging low-energy component in the distributions. Advection of particles from the upstream towards the plasmoids, where they lose their bulk motion energy, could explain this behavior, although only an analysis of particle trajectories could validate such a hypothesis. 
In summary, the presence of a shear flow decreases the particle acceleration efficiency: we observe a clear trend towards a softening of the particle energy distribution with increasing shear.

\section{Conclusions}\label{sec:Conclusion}

In this work, we analyze separately the impacts of composition asymmetry and shear in the upstream plasma of a relativistic reconnecting current layer. For composition-asymmetric reconnection, the less-magnetized electron-ion upstream side dictates the global reconnection dynamics, in agreement with \citet{Mbarek2022}. The measured reconnection rate is very close to, although slightly higher than, that reported in previous symmetric electron-ion reconnection studies \citep{Werner2018, Ball2018}. Furthermore, we notice that the gap in the energy partition between electrons and ions increases by about 40\% with respect to a symmetric electron-ion configuration, meaning that ions are even more preferentially heated in the presence of composition asymmetry. In the second part of our work, focused on a sheared reconnection layer, we show that a relativistic super-Alfv\'enic shear substantially reduces the reconnection rate but, unlike in non-relativistic studies, does not completely halt reconnection. In addition, we observe a steepening of the power-law tail for increasing shear values.

These results have important implications for astrophysical environments. In the context of black hole accretion disks, asymmetric reconnection may occur at a jet-disk interface or during magnetic flux eruptions. One important asymmetry in such reconnection events is with respect to plasma composition. In this case, our results indicate that particle acceleration is dictated by the heavier -- and, hence, lower-magnetization -- electron-ion side of the current sheet. Due to the modest ion magnetization ($\sigma\lesssim 1$) expected in this environment, reconnection is unlikely to produce prominent power-law tails in the ion energy distribution; a prominent nonthermal component exceeding $\gamma = m_i/m_e \sigma \lesssim m_i/m_e$ in the electron energy distribution is also not expected.

These properties suggest that asymmetric reconnection at a jet-disk boundary is not a suitable candidate for powering high-energy gamma-ray flares observed in black hole environments. On the other hand, such reconnection sites may contribute to the lower-energy emission observed in limb-brightened jets. The addition of a velocity shear along field lines at the jet-disk boundary does not change these expectations, since, as we show in this work, strong shear slows reconnection down and inhibits nonthermal particle acceleration -- effects that would only further hamper efficient acceleration of particles up to gamma-ray-emitting energies. 

An important limitation of our study is its dimensionality. Once particles are trapped in two-dimensional plasmoids, they are artificially kept from escaping back into the upstream plasma. However, in 3D, as shown by recent studies \citep{Zhang2021, 2023ApJ...959..122C}, ions can escape plasmoids and get onto Speiser-like trajectories where they undergo intense linear acceleration. In the presence of composition asymmetry, this mechanism may facilitate mixing of ions into the putatively Poynting-flux-dominated and pair-loaded jet. It may also facilitate particle injection into other acceleration processes occurring on larger scales, such as, for instance, shear-flow acceleration along the jet sheath \citep{1998A&A...335..134O, 2004ApJ...617..155R, Sironi2021}. This motivates future studies of 3D composition-asymmetric relativistic magnetic reconnection.

\begin{acknowledgements}
We thank Jesse Vos for providing us early results from his GRMHD simulations and for fruitful discussions regarding magnetized black-hole accretion. We are thankful to the referee's comments that helped us improving the quality of the manuscript. This project has received funding from the European Research Council (ERC) under the European Union’s Horizon 2020 research and innovation program (Grant Agreement No. 863412). Computing resources were provided by TGCC under the allocation A0150407669 made by GENCI.
\end{acknowledgements}

%
%

 \bibliographystyle{aa}
 \bibliography{bibliograpy}

\end{document}